\documentclass[a4paper,12pt]{article}
\usepackage{amsfonts,amsmath,epsfig,subfigure} 
\begin{document} 
\numberwithin{equation}{section} 
\title{Explicit secular equation for Scholte waves \\
       over a monoclinic crystal.} 

\author{Michel Destrade}
\date{2004} 
\maketitle 

\section{Introduction} 

Scholte waves are acoustic waves propagating at a fluid/solid 
interface. 
They are localized in the neighborhood of the phase boundary in the 
sense that they decay exponentially in both directions along the 
normal to the interface.
Johnson \cite{John70} established the explicit secular equation for 
Scholte waves over an orthorhombic crystal. 
In his case, the crystal is cut along a plane $x_2=0$ containing 
\textit{two} crystallographic axes $Ox_1$ and $Ox_3$; 
the wave propagates with speed $v$ in the $x_1$ direction;
the solid is characterized by a mass density $\rho_s$ and  relevant 
elastic stiffnesses $C_{11}$,  $C_{12}$,  $C_{22}$, and $C_{66}$; 
the fluid by a  mass density $\rho_f$ and speed of sound $c$. 
The secular equation is 
\begin{align} \label{seculOrtho}
& Z \sqrt{C_{11}C_{22} - C_{12}^2 - 2C_{12}C_{66} - (C_{22}+C_{66})X 
           + 2\sqrt{C_{22}C_{66}(C_{11}-X)(C_{66} - X)}} 
\nonumber \\ 
& \phantom{1234567890} - \sqrt{\frac{C_{66} - X}{C_{11} - X}} 
           (C_{11}C_{22} - C_{12}^2 - C_{22} X) 
                 + X \sqrt{C_{22}C_{66}} = 0, 
\end{align} 
where
\begin{equation} \label{Z}
X = \rho_s v^2, \quad 
Z = \frac{\rho_f v^2}{\sqrt{1 - \frac{v^2}{c^2}}}. 
\end{equation} 
For instance, consider a frozen lake with a layer of ice assumed thick 
enough to be considered as a semi-infinite body. 
At 0.01$^\text{o}$C under 1 bar, the density of water is 
\cite{LeMF03}: $\rho_f = 999.84$ kg/m$^3$ and sound propagates at 
$c = 1402.4$ m/s; 
the second line of Table 1 lists the elastic stiffnesses and density 
of ice \cite{Shut88}; 
according to \eqref{seculOrtho}, Scholte waves propagate for this 
model at speed $v_S = 1237.6$ m/s.
Ice however has the special property of being transversally isotropic, 
which means that any plane containing the $x_3$ axis is a symmetry 
plane and so the speed $v_S$ is the same for any orientation of the 
water/ice interface plane containing the $x_3$ axis.

The aim of this Letter to the Editor is to derive explicitly the 
secular equation for Scholte waves at the interface between a fluid 
and an anisotropic crystal cut along a plane containing the normal to 
a single symmetry plane, that is containing only \textit{one} 
crystallographic axis. 
In effect, the crystal may be a monoclinic crystal with symmetry plane 
at $x_3=0$, or a rhombic, tetragonal, or cubic crystal cut along a 
plane containing $x_3$ and making an angle $\theta \ne 0$ with the 
other crystallographic planes; 
the higher symmetry cases ($\theta = 0$ or transversally isotropic 
and isotropic crystals) are covered by \eqref{seculOrtho}.
For cases with less symmetries, one can turn to approximate solutions
\cite{NoSi95} as long as the anisotropy is weak .

\section{Equations of motion and boundary conditions} 
 
Consider two half-spaces delimited by the plane $x_2=0$; 
the upper one $x_2<0$ is filled with an inviscid fluid, 
the lower one $x_2>0$ is made of a monoclinic crystal with symmetry 
plane at $x_3=0$ whose relevant non-zero reduced compliances are 
$s'_{11}$, $s'_{22}$, $s'_{12}$, $s'_{16}$, $s'_{26}$, and $s'_{66}$. 
At the interface, an inhomogeneous plane wave travels with speed $v$ 
and wave number $k$ in the $x_1$ direction, and decays rapidly in the 
$x_2 \rightarrow \pm \infty$ directions. 

In the \textit{solid}, the corresponding equations of motion are 
written as a first-order differential system for the 4-component 
displacement-traction vector, 
\begin{equation}  \label{motion}
\mbox{\boldmath $\xi$}'= \text{i}\mathbf{N}\mbox{\boldmath $\xi$}, 
\quad 
\mbox{\boldmath $\xi$}(kx_2) = 
 [U_1(kx_2), U_2(kx_2), t_{12}(kx_2), t_{22}(kx_2)]^\text{T}, 
\end{equation}
where the functions $U_i$ and $t_{i2}$ are related to the in-plane 
mechanical displacements $u_1$, $u_2$ and in-plane 
tractions $\sigma_{12}$, $\sigma_{22}$ through 
\begin{equation} \label{wave} 
u_i(x_1, x_2, x_3, t) = U_i(kx_2)\text{e}^{\text{i}k(x_1 - vt)}, 
\quad 
\sigma_{i2} (x_1, x_2, x_3, t) 
    = \text{i}kt_{i2}(kx_2)\text{e}^{\text{i}k(x_1 - vt)}.
\end{equation}
In \eqref{motion}, the $4 \times 4$ matrix $\mathbf{N}$ is given by 
\cite{Dest01, Ting02}, 
\begin{equation}
\mathbf{N} =
 \begin{bmatrix} 
  -r_6 &  - 1  & n_{66}  &  n_{26}  \\
  -r_2 &    0  & n_{26}  &  n_{66}  \\
 X-\eta&    0  &  -r_6   &  -r_2    \\  
   0   &    X  &  - 1    &    0   
 \end{bmatrix}, 
\end{equation}
where $X=\rho_s v^2$ and 
\begin{equation} 
 \eta =   \frac{1}{s'_{11}},
\quad 
r_i  =  -\frac{s'_{1i}}{s'_{11}},
\quad
n_{ij} =   \frac{1}{s'_{11}}\begin{vmatrix}
                                         s'_{11}  & s'_{1j} \\
                                         s'_{1i}  & s'_{ij}
                              \end{vmatrix}. 
\end{equation}
These equations also cover the case of a wave \eqref{wave} travelling 
in a crystal of rhombic,  tetragonal, or cubic symmetry, 
with acoustic axes $XYx_3$ and reduced compliances $S'_{ij}$, 
cut along the plane $x_2=0$ containing the $x_3$ axis and making an 
angle $\theta$ with the crystallographic $XY$ plane  (see Figure 1).
In that case, the reduced compliances $s'_{ij}$ along the $x_i$ axes 
are given in terms of those along the crystallographic axes $XYx_3$ by 
(see  Ting \cite{Ting00}), 
\begin{align} \label{s'rotated}
& s'_{11} = S'_{11}\cos^4 \theta 
            + (2S'_{12} + S'_{66})\cos^2 \theta \sin^2 \theta 
             + S'_{22}\sin^4 \theta,
\nonumber \\
& s'_{22} = S'_{22}\cos^4 \theta 
            + (2S'_{12} + S'_{66})\cos^2 \theta \sin^2 \theta
             + S'_{11}\sin^4 \theta,
\nonumber \\
& s'_{12} = 
  S'_{12} 
  + (S'_{11}+S'_{22}-2S'_{12}-S'_{66})\cos^2 \theta \sin^2 \theta,
\nonumber \\
& s'_{66} = 
S'_{66}
  + 4(S'_{11}+S'_{22}-2S'_{12}-S'_{66})\cos^2 \theta \sin^2 \theta.
\nonumber \\
& s'_{16} = 
  [2S'_{22}\sin^2 \theta  - 2S'_{11}\cos^2 \theta
   + (2S'_{12} + S'_{66})(\cos^2 \theta - \sin^2 \theta)]
                                             \cos \theta \sin \theta,
\nonumber \\
& s'_{26} = 
  [2S'_{22}\cos^2 \theta  - 2S'_{11}\sin^2 \theta
    - (2S'_{12} + S'_{66})(\cos^2 \theta - \sin^2 \theta)]
                                             \cos \theta \sin \theta.
\end{align}
Note that for transversally isotropic crystals, the following 
relationships hold, $S'_{11} = S'_{22}$, 
$S'_{66} = 2(S'_{11} - S'_{12})$, and the rotation does not affect the 
values of the compliances ($s'_{ij} = S'_{ij}$). 
This author \cite{Dest03} recently showed that for waves vanishing 
with increasing distance from the plane $x_2=0$, the following 
fundamental relationships hold for any positive or negative integer 
power $n$ of the matrix $\mathbf{N}$,
\begin{equation} \label{equations}
\overline{\mbox{\boldmath $\xi$}}(0)  \cdot 
   \widehat{\mathbf{I}}\mathbf{N}^n \mbox{\boldmath $\xi$}(0) =  0, 
\quad \text{where} \quad 
 \widehat{\mathbf{I}} = 
  \begin{bmatrix} 0 & 0 & 1 & 0 \\
                  0 & 0 & 0 & 1 \\ 
                  1 & 0 & 0 & 0 \\
                  0 & 1 & 0 & 0 
  \end{bmatrix}.
\end{equation}
Because of the Cayley-Hamilton theorem, only three consecutive 
powers of $\mathbf{N}$ are linearly independent so that 
\eqref{equations} reduces to only three linearly independent 
equations. 

In the \textit{fluid}, the normal displacement and the normal stress 
component are connected, as recalled by Barnett et al. \cite{BaGL88}, 
by the (real) normal impedance $Z$ defined in \eqref{Z}$_2$, 
\begin{equation} \label{fluid} 
\sigma_{22} = k Z u_2. 
\end{equation}

At the \textit{solid/fluid interface}, the normal displacement and the 
normal stress component are continuous, and the shear stress component 
is zero. 
It follows from these boundary conditions and from \eqref{motion}$_2$, 
\eqref{wave}, \eqref{fluid}, that the displacement-traction vector 
at the interface $x_2 = 0^+$ is of the form, 
\begin{equation}  \label{interface}
\mbox{\boldmath $\xi$}(0^+) = 
 U_2(0)[\alpha, 1, 0, -\text{i}Z]^\text{T}, 
\end{equation}
where $\alpha = U_1(0^+)/U_2(0)$.

Now the fundamental equations \eqref{equations} read
\begin{equation} 
(N^n)_{32}(\alpha + \overline{\alpha})
     + \text{i} Z (N^n)_{21}(\alpha - \overline{\alpha}) 
      + (N^n)_{31} \alpha \overline{\alpha}
         = - (N^n)_{42} - Z^2 (N^n)_{24}.
\end{equation}
Writing $\alpha$ as $\alpha = \alpha_1 + \text{i} \alpha_2$ and taking 
in turn $n = -1, 1, 2$, a non-homogeneous linear system of 
equations follows, 
\begin{multline} \label{system}  
\mathbf{A  b =d}, 
\quad 
\mathbf{A} = 
\begin{bmatrix}
  N^*_{32}   & Z N^*_{22}  & N^*_{31}   \\ 
     0       & Z N_{22}    & N_{31}     \\ 
 (N^2)_{32}  & Z (N^2)_{22}& (N^2)_{31} 
\end{bmatrix},
\\
\mathbf{b} = 
\begin{bmatrix}
   2 \alpha_1 \\ -2\alpha_2 \\ \alpha_1^2+ \alpha_2^2 
\end{bmatrix},
\quad
\mathbf{d} = 
 - \begin{bmatrix}
   N^*_{42} + Z^2 N^*_{24}   \\     
   N_{42} + Z^2 N_{24}       \\
   Z^2 (N^2)_{24}
\end{bmatrix},
\end{multline}
where $\mathbf{N^*}$ denotes the adjoint of $\mathbf{N}$.
The unique solutions to the system are $b_k = \Delta_k/\Delta$, 
where $\Delta = \text{ det } \mathbf{A}$ and $\Delta_k$ is the 
determinant of the matrix derived from $\mathbf{A}$ by replacing the 
$k$-th column with $\mathbf{d}$.
However, the $b_k$ are linked by $b_1^2+b_2^2 = 4 b_3$, which is 
the explicit secular equation for Scholte wave over a monoclinic 
crystal with symmetry plane at $x_3 = 0$, 
\begin{equation} \label{seculScholte}
\Delta_1^2 + \Delta_2^2 = 4 \Delta \Delta_3. 
\end{equation}

As a check, the limit case of a solid/vacuum interface is examined. 
When the density of the fluid $\rho_f$ is taken as zero, then 
by \eqref{Z}$_2$ $Z = 0$, and so $\Delta = \Delta_1 = \Delta_3 = 0$. 
The secular equation reduces to $\Delta_2 = 0$ (written at $Z=0$), 
that is the following quartic in $X = \rho_s v^2$ 
\cite{Curr79, Dest01, Ting02}, 
\begin{equation}  \label{seculSAW}
\begin{vmatrix} 
X[r_2r_6 - n_{26}(X-\eta)]  
                & (X-\eta)(1+n_{66}X)+r_6^2X 
                           &  X[r_2^2 - n_{66}(X-\eta)] \\
         0      &     X    &         X-\eta             \\
 (1+r_2)X- \eta &     0    &      2r_6(X-\eta) 
\end{vmatrix} = 0.
\end{equation}

\section{Examples} 

Calculations for usual combinations of a solid and a fluid show that 
in general the speed of a Scholte wave is very close to the speed of 
sound in the fluid. 
Hence, consider  water ($\rho_f = 1025$ kg/m$^3$, $c = 1531$ m/s 
at 25$^\text{o}$C \cite{Weas71}) over gypsum (monoclinic, $\rho_s$ 
and $C_{ij}$ in Table 1 \cite{ChWi92}): 
the secular equation \eqref{seculScholte} yields a Scholte wave speed 
within the interval [1519 m/s, 1526 m/s] (depending on the orientation 
of the cut plane), which is within less than 0.8\% of the speed of 
sound in the water and beyond reasonable accuracy for measurements.

Yet for certain choices, the Scholte wave speed moves away 
from the speed of sound in the fluid. 
One example is the combination ice/water presented in the 
Introduction. 
A second example is the combination of pure water 
($\rho_f = 998$ kg/m$^3$, $c = 1498$ m/s at 25$^\text{o}$C 
\cite{Weas71}) and Terpine Monohydride  
(orthorhombic, $\rho_s$ and $C_{ij}$ in Table 1 \cite{Shut88}):
at $\theta=0^\text{o}$ and $\theta=90^\text{o}$ (crystal cut along a 
plane containing two crystallographic axes) the wave propagates at 
1228.3 m/s and 1249.5 m/s, respectively; 
Figure 2(a) shows how the Scholte wave speed varies between these two 
extremes as a function of $\theta$. 
Another way of separating distinctly the Scholte wave speed from the 
sound speed is to increase the pressure, and hence the speed of sound, 
in the fluid. 
Crowhurst  \cite{CASB01} et al. recently measured the Scholte wave 
speed for Methanol over Germanium in a diamond anvil cell: 
as the pressure increases from 0.56 GPa to 2.2 GPa, so does the 
speed of sound in Methanol, from about 2500 m/s to 3500 m/s. 
In Table 1, the stiffnesses and density of Germanium (cubic) at 
$20^\text{o}$ are recalled \cite{Shut88}; 
the density of Methanol is 791.4 kg/m$^3$ at $20^\text{o}$ 
\cite{Weas71}. 
Figure 2(b) shows, in agreement with their results, the combined 
influence of orientation and speed of sound on Scholte wave 
propagation; 
each curve corresponds to a different speed of sound in Methanol, 
from $c = 2000$ m/s (bottom curve) to $c = 4000$ m/s (top curve) by 
500 m/s increments.


        
\newpage

\begin{figure}
\centering
\mbox{\epsfig{figure=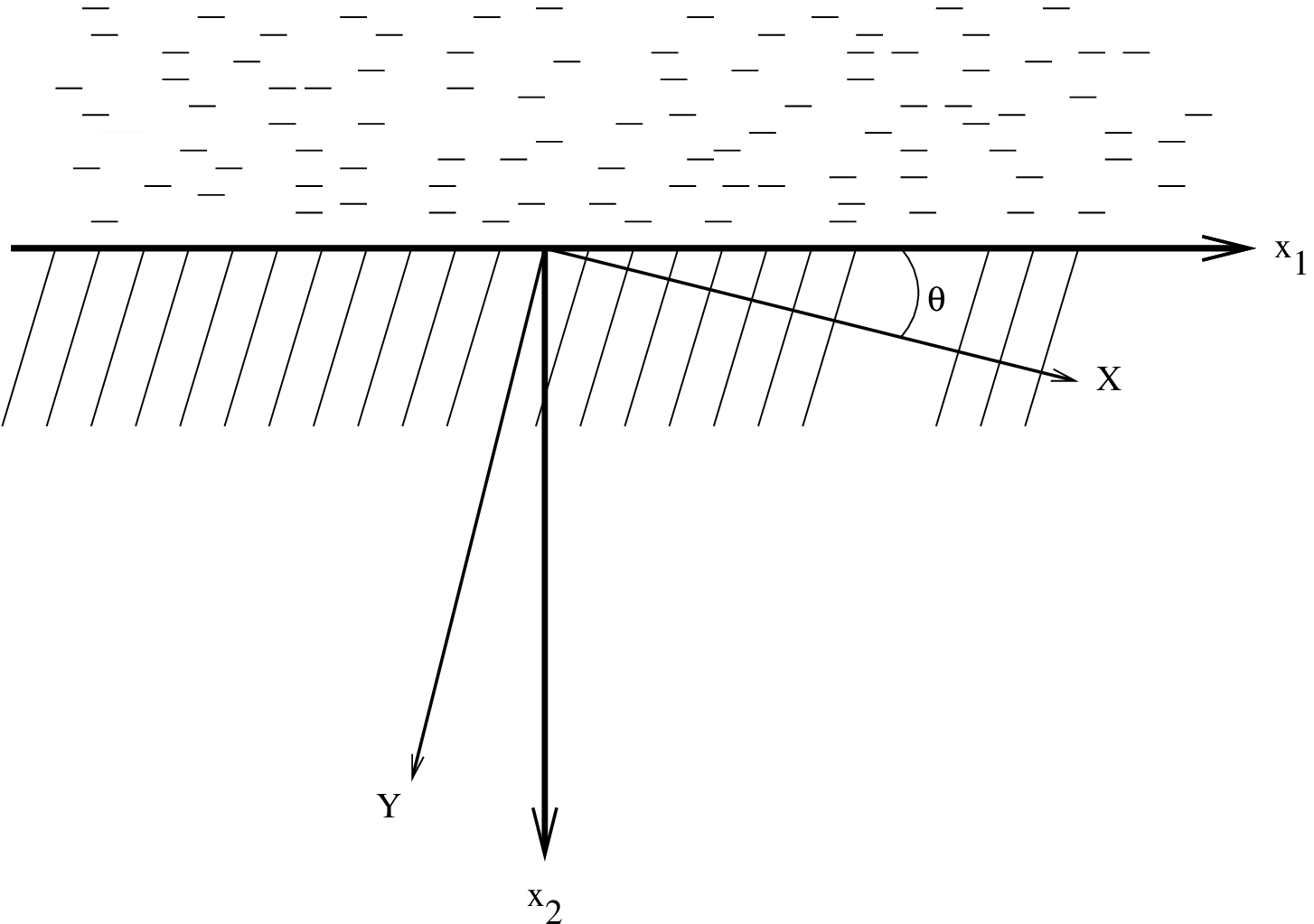, width=.8\textwidth}}
 \caption{Fluid/solid interface}
\end{figure}
  
\vspace*{\fill}
     
\newpage

\begin{figure}
\centering
\mbox{\subfigure{\epsfig{figure=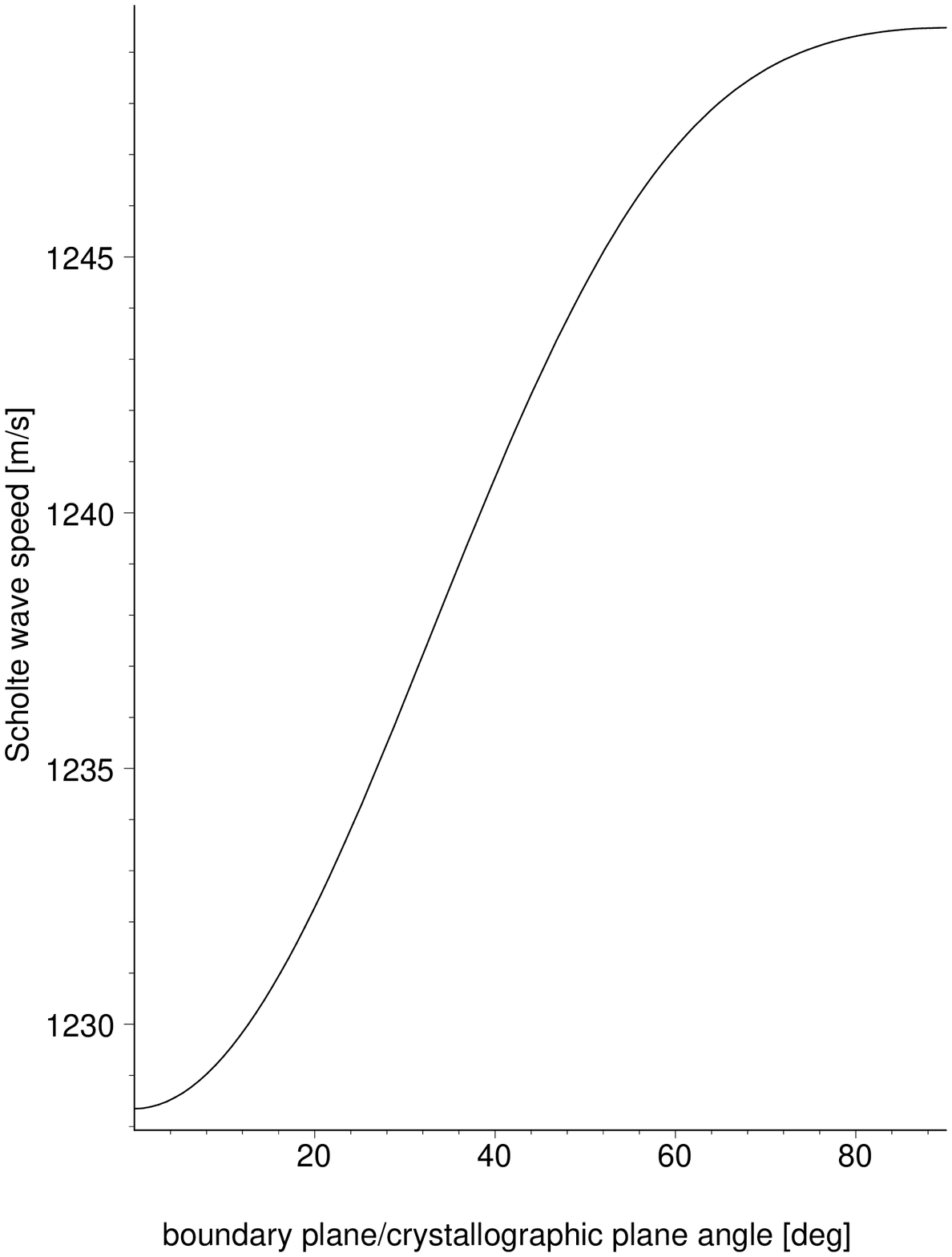,
width=.45\textwidth}}
  \quad \quad
     \subfigure{\epsfig{figure=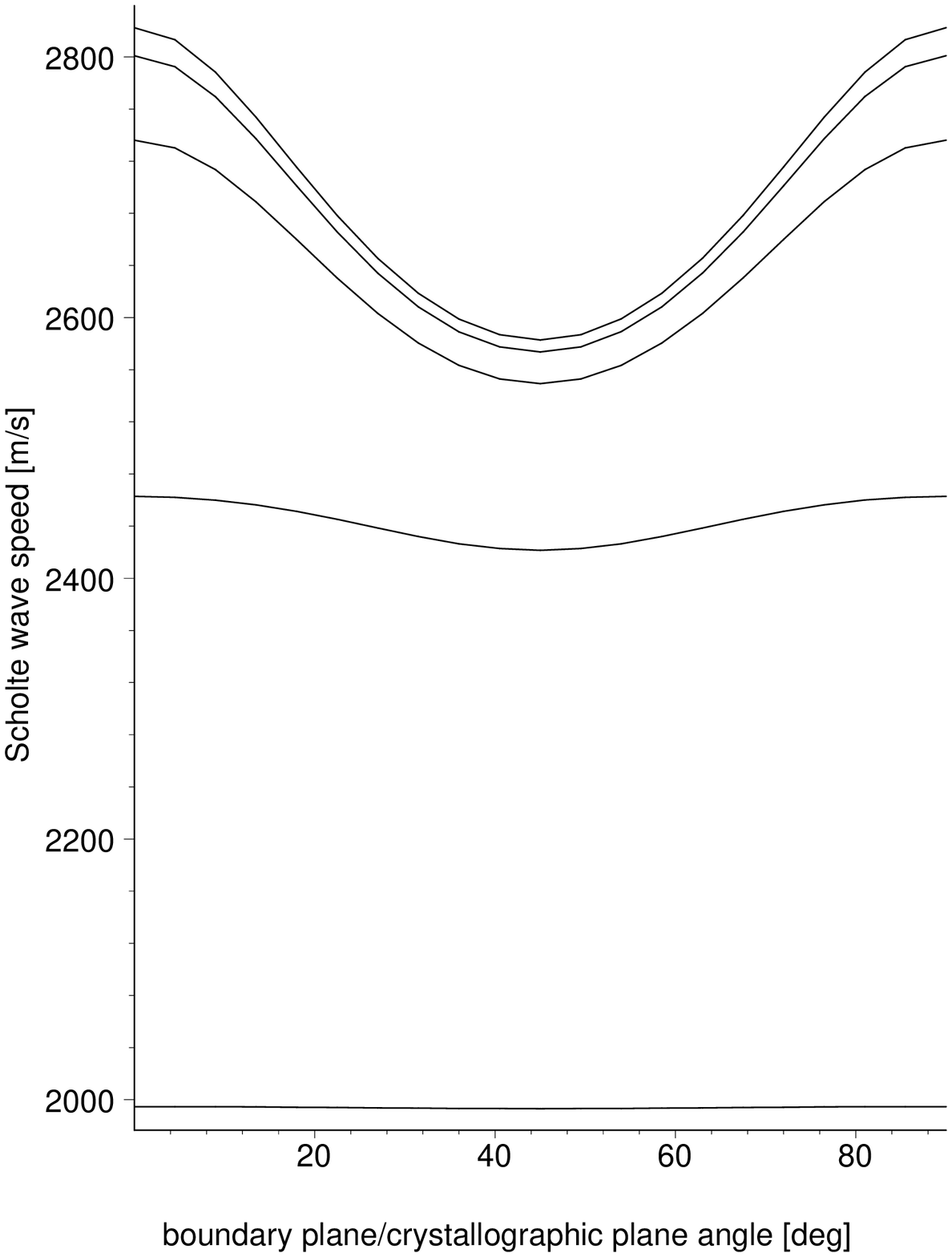,
width=.45\textwidth}}}
\caption{Scholte wave speeds for (a) Water/Terpine interface and
 (b) Methanol/Germanium interface, where the speed of sound in the 
fluid is [m/s]: 2000 (bottom curve), 2500, 3000, 3500, 4000 
(top curve).}
\end{figure}
 
\vspace*{\fill}

\newpage

\bigskip

\noindent

{\Large\textbf{List of Figures.}}

\bigskip
\noindent
\textbf{Figure 1: Fluid/solid interface.}

\bigskip
\noindent
\textbf{Figure 2: Scholte wave speeds (a) Water/Terpine interface and
 (b) Methanol/Germanium interface, where the speed of sound in the 
fluid is [m/s]: 2000 (bottom curve), 2500, 3000, 3500, 4000 
(top curve).}

\textbf{Figure 2(a):}

\noindent
Legend on graduated horizontal axes: ``boundary plane/crystallographic 
plane angle [deg].''

\noindent
Legend on graduated vertical axis: ``Scholte wave speed [m/s].''

\medskip

\textbf{Figure 2(b):}

\noindent
Legend on graduated horizontal axes: ``boundary plane/crystallographic 
plane angle [deg].''

\noindent
Legend on graduated vertical axis: ``Scholte wave speed [m/s].''


\newpage

\begin{center}
Table 1.
\textit{Values of the elastic stiffnesses} ($10^{10}$ N/m$^2$), 
\textit{density} (kg/m$^3$),
\textit{and surface (Rayleigh) wave speed} (m/s) 
\textit{for 3 crystals.}

\noindent
\begin{tabular}{l c c c c c c c c}
\hline
\rule[-3mm]{0mm}{8mm} 
crystal & $C_{11}$  & $C_{22}$     & $C_{12}$    & $C_{16}$
 & $C_{26}$      & $C_{66}$    & $\rho_s$     & $v_R$
\\
\hline
ice ($-5^\text{o}$C)             
                & 1.38 & 1.38 & 0.707 &   0   &   0   & 0.3365
                & 940  & 1766
\\
gypsum          & 50.2 & 94.5 & 28.2  & -7.5  & -11.0 & 32.4 
                & 2310 & 3011
\\
terpine         & 1.25 & 0.99 & 0.38  &   0   &   0   & 0.346 
                & 1110 & 1644
\\
germanium       & 12.92& 12.92& 4.79  &   0   &   0   & 6.70 
                & 5320 & 2936
\\
 \hline
\end{tabular}
\end{center}

\end{document}